\documentclass[10pt,twocolumn,preprintnumbers,amsmath,amssymb,nofootinbib,superscriptaddress,aps,prd,numberedappendix]{openjournal}

\usepackage{newtxtext,newtxmath}
\usepackage{lipsum}
\usepackage[T1]{fontenc}
\usepackage{ae,aecompl}

\usepackage{graphicx}	
\usepackage{amsmath}	
\usepackage{amssymb}	

\newcommand{\henk}[1]{#1}
\newcommand{\extra}[1]{#1}
\newcommand{\peter}[1]{#1}

\begin{document}

\title[Propagating  Biases in Cosmic Shear Power Spectra]{Propagating \henk{Residual} Biases in Cosmic Shear Power Spectra}

\author{T. D. Kitching}
\email{t.kitching@ucl.ac.uk}
\affiliation{Mullard Space Science Laboratory, University College London, Holmbury St Mary, Dorking, Surrey RH5 6NT, UK}

\author{P. Paykari}
\affiliation{Mullard Space Science Laboratory, University College London, Holmbury St Mary, Dorking, Surrey RH5 6NT, UK}

\author{H. Hoekstra}
\affiliation{Leiden Observatory, Leiden University, Niels Bohrweg 2, NL-2333 CA Leiden, The Netherlands}

\author{M. Cropper}
\affiliation{Mullard Space Science Laboratory, University College London, Holmbury St Mary, Dorking, Surrey RH5 6NT, UK}

\begin{abstract}
In this paper we derive a full expression for the propagation of multiplicative and additive shape measurement biases into the cosmic shear power spectrum. In doing so we identify several new terms that are associated with selection effects, as well as cross-correlation terms between the multiplicative and additive biases and the shear field. \extra{The computation of the resulting bias in the shear power spectrum scales as the fifth power of the maximum multipole considered. Consequently the  
calculation is unfeasible for large $\ell$-modes, and the only tractable way to assess the full impact of shape measurement biases on cosmic shear power spectrum is through forward modelling of the effects. To linear order in bias parameters the shear power spectrum is only affected by the mean of the multiplicative bias field over a survey and the cross correlation between the additive bias field and the shear field. If the mean multiplicative bias is zero then second order convolutive terms are expected to be orders of magnitude smaller.}  
\end{abstract}

\maketitle

\section{Introduction}
\label{S:Intro}

The statistical properties of the large-scale matter distribution over cosmic time encodes key information about the late time evolution of the Universe, and also allows us to improve constraints on the initial conditions. Thanks to technological advances we can now efficiently survey larger and larger areas of sky, but the interpretation of galaxy redshift surveys \henk{is} hampered by the fact that \henk{galaxies} are biased tracers of the underlying dark matter distribution. Fortunately, the distortion of space-time by matter results in correlations in the ellipticities of distant galaxies that are the result of the differential deflection of light rays, a phenomenon called gravitational lensing. \henk{The statistics of these correlations can be directly related to those of the large-scale structure.} This in turn enables us to constrain the nature of dark energy and to test gravity on cosmological scales.

The cosmological lensing signal has now been robustly measured using large ground-based imaging surveys \citep[e.g.][]{KV, des}. However to shed light on the nature of dark energy, the precision needs to increase significantly. This is the objective of a number of planned projects that \henk{will commence soon}. {\it Euclid} \citep{redbook} aims to image 15\,000 deg$^2$ of extragalactic sky from space, while the Large Synoptic Survey Telescope (LSST) will survey a similar area from the ground. To exploit fully the potential of these data for cosmology, it is essential that astrophysical and instrumental sources of biases are accounted for at levels that are small compared to the statistical uncertainties on measured cosmological parameters. Accurate measurements of the shapes of small, faint galaxies are therefore essential.

\henk{The observed ellipticities of galaxies used in weak lensing studies are typically biased} with respect to the true ellipticities that would have been measured given ideal data and an ideal measurement algorithm. The dominant sources of bias are a result of the \henk{convolution by the} point spread function (PSF) and noise in the images. For this reason the performance of shape measurements has been studied extensively \citep{step1,step2,great08,great10,great3}.

To first order the biases can be separated into multiplicative and additive functions that act on the true shear. Additive biases arise from anisotropies in the data, such as an anisotropic PSF or detector effects. These do not only affect the measurement of the galaxy shape, but also the detection and selection of sources. Characterizing and correcting for these sources of bias is essential, but residual spurious alignments might still be removed through empirical corrections. For instance, the mean shear when averaged in the coordinate frame defined by the detector should vanish. The detection of a coherent signal would thus indicate an imperfect correction, but that signal could also be fitted for in the cosmological analysis. However we note that this is only partially effective because sources of additive bias are expected to introduce multiplicative biases with a similar amplitude. Unfortunately, multiplicative bias cannot be readily inferred from the imaging data directly. Instead image simulations are used to calibrate the biases in the shape measurement algorithms \citep[e.g.][]{cccp, 2018arXiv181203983K}, although we note that alternative approaches have been recently proposed \citep{2017arXiv170202600H}. 

The desired accuracy in cosmological parameter estimates determines the level at which shape measurement biases can be tolerated. The propagation of biases, or residual biases after calibration, into the weak lensing power spectra (or `cosmic shear' power spectrum) is not straightforward. Some studies approximate the full expression \citep{tk,great10,massey} but, as we show in this paper, these results do not capture the spatially varying sources of biases correctly. This is of particular importance because the theoretical propagation of such biases into power spectrum residuals drives the design requirements for experiments that use weak lensing as a cosmological probe \citep{cropper}. 

In this paper we show how multiplicative and additive biases in shape measurement propagate through the cosmic shear power spectra, discuss how this formalism relates to previous studies, and discuss the implications for the assessment of shape measurement biases on cosmological parameter performance verification of experiments. In Section \ref{S:Method} we present the formalism, in Section \ref{Simple Simulations} we present some simple simulations that demonstrate the accuracy of the formalism, and in Section \ref{Discussion} we examine the implications of this study; conclusions are presented in Section \ref{Conclusions}. 

\section{Method}
\label{S:Method}
We begin with the expression for the measured shear in real (angular) space 
\begin{equation}
    \label{gamma}
    \widetilde\gamma(\mathbf{\Omega})=[1+m_0+m(\mathbf{\Omega})]\gamma(\mathbf{\Omega})+[c_{1,0}+{\rm i}c_{2,0}+c(\mathbf{\Omega})],
\end{equation}
where $\widetilde\gamma(\mathbf{\Omega})$ is the measured shear as a function of angle $\mathbf{\Omega}=(\theta,\phi)$ where $\theta$ and $\phi$ are arbitrary spherical coordinates, $m_0$ is a constant multiplicative bias, $m(\mathbf{\Omega})=m^R(\mathbf{\Omega})+{\rm i}m^I(\mathbf{\Omega})$ is a position-dependent multiplicative bias term, $\gamma(\mathbf{\Omega})=\gamma_1(\mathbf{\Omega})+{\rm i}\gamma_2(\mathbf{\Omega})$ is the true shear, $c_{1,0}$ and $c_{2,0}$ are constant additive biases \peter{that contribute to the real and imaginary parts of the additive field}, and $c(\mathbf{\Omega})$ is a position-dependent additive bias. We assume no non-local terms, e.g. $m(\mathbf{\Omega}')\gamma(\mathbf{\Omega})$, since such terms could be always re-written as a local per galaxy $m(\mathbf{\Omega})\gamma(\mathbf{\Omega})$ term. 

\peter{We discuss the choice of the multiplicative bias expression in Appendix A. We note that the choice to express the multiplicative effect as a product of complex numbers makes the spin-preserving assumption that any effect only changes the amplitude and/or angle of the observed ellipse relative to the unbiased case. There are more general expressions that can be used to capture biases that can occur in the case of anisotropic systematic effects, however image simulations suggest that our adopted approach is accurate for residual systematic effects after calibration with image simulations. In the case where no rotational change is present this reduces to a multiplication by a single scalar field $m(\mathbf{\Omega})$. We will revisit these assumptions later in the analysis.}

\subsection{Spherical Harmonic Representation}
We now determine the spherical harmonic representation of a biased shear field. We adapt the methodology from CMB pseudo-$C_{\ell}$ analysis here for the general bias case; in particular we follow \cite{lct,bct,zs,grain} but we generalize their formalism further to include general spin-2 bias functions and additive terms. In \cite{great10} a similar adaption was made, but under simplifying assumptions that did not capture the general case. 

Since for cosmic shear the cosmological information is contained within the E-mode (gradient) component of the field, and not in the B-mode (curl) component, we work on spherical harmonic coefficients $\gamma^E_{\ell m}$ and $\gamma^B_{\ell m}$. We can write down the E- and B-mode coefficients in terms of the shear field as 
\begin{eqnarray}
    \label{gtoe}
    \gamma^E_{\ell m}&=&\frac{1}{2}\int {\rm d}\mathbf{\Omega}\, [
    \gamma(\mathbf{\Omega})\,{}_2Y^*_{\ell m}(\mathbf{\Omega})+
    \gamma^*(\mathbf{\Omega})\,{}_{-2}Y^*_{\ell m}(\mathbf{\Omega})]\nonumber\\
    \gamma^B_{\ell m}&=&\frac{-{\rm i}}{2}\int {\rm d}\mathbf{\Omega}\, [
    \gamma(\mathbf{\Omega})\,{}_2Y^*_{\ell m}(\mathbf{\Omega})-
    \gamma^*(\mathbf{\Omega})\,{}_{-2}Y^*_{\ell m}(\mathbf{\Omega})], 
\end{eqnarray}
where $\ell$ and $m$ are angular wavenumbers\footnote{Note that $m$ is used in the spherical harmonic function, and $m_0$ and $m(\mathbf{\Omega})$ as multiplicative biases; we choose to keep this standard notation for both cases as the use should be clear from the context.}, ${}_2Y_{\ell m}(\mathbf{\Omega})$ is the standard spin-weighted spherical harmonic function for a spin-$2$ field and $^*$ denotes a complex conjugate. This expression is exact for an all-sky unbiased measurement of the shear. This formalism could be generalised to the pure-mode case \citep{grain}, that would become important in the presence of masks, but we leave this masked data generalisation for future work.  

To compute the effect of the biases we now replace $\gamma(\mathbf{\Omega})$ in equation (\ref{gtoe}) with $\widetilde\gamma(\mathbf{\Omega})$ from equation (\ref{gamma}). This results in the following expressions
\begin{eqnarray}
    \label{shtransform}
    \widetilde\gamma^E_{\ell m}&=&
    \gamma^E_{\ell m}+
    m_0\gamma^E_{\ell m}\nonumber\\
    &+&\sum_{\ell'm'}
    [\gamma^E_{\ell'm'}W^+_{\ell\ell'm m'}+
     \gamma^B_{\ell'm'}W^-_{\ell\ell'm m'}]\nonumber\\
    &+&c^E_{\ell m}\nonumber\\
    \widetilde\gamma^B_{\ell m}&=&
    \gamma^B_{\ell m}+
    m_0\gamma^B_{\ell m}\nonumber\\
    &+&\sum_{\ell'm'}
    [\gamma^B_{\ell'm'}W^+_{\ell\ell'm m'}-
     \gamma^E_{\ell'm'}W^-_{\ell\ell'm m'}]\nonumber\\
    &+&c^B_{\ell m}.
\end{eqnarray}

\noindent \henk{Here we expanded the additive term as $c(\mathbf{\Omega})=\sum_{\ell m}(c^E_{\ell m}+{\rm i}c^B_{\ell m})_{2}Y_{\ell m}(\mathbf{\Omega})$. We also defined}
\begin{eqnarray}
    W^+_{\ell\ell' m m'}&=&\frac{1}{2}[
    (_{2}W^{R,mm'}_{\ell\ell'}+_{-2}W^{R,mm'}_{\ell\ell'})\nonumber\\&+&
    {\rm i}(_{2}W^{I,mm'}_{\ell\ell'}-_{-2}W^{I,mm'}_{\ell\ell'})],\nonumber\\
    W^-_{\ell\ell' m m'}&=&\frac{{\rm i}}{2}[
    (_{2}W^{R,mm'}_{\ell\ell'}-_{-2}W^{R,mm'}_{\ell\ell'})\nonumber\\&+&
    {\rm i}(_{2}W^{I,mm'}_{\ell\ell'}+_{-2}W^{I,mm'}_{\ell\ell'})],
\end{eqnarray}
and 
\begin{eqnarray}
    _{s}W^{R,mm'}_{\ell\ell'}=\int{\rm d}\mathbf{\Omega}\, _{s}Y^*_{\ell'm'}(\mathbf{\Omega})m^R(\mathbf{\Omega})_{s}Y_{\ell m}(\mathbf{\Omega});
\end{eqnarray}
and similarly for $m^I(\mathbf{\Omega})$.
\henk{In this derivation} we have expressed the real and imaginary parts of the multiplicative bias as $m(\mathbf{\Omega})=m^R(\mathbf{\Omega})+{\rm i}m^I(\mathbf{\Omega})$. \peter{We note that when considering \emph{residual} systematic effects, i.e. when any amplitude and rotational changes caused by multiplicative systematic effects are small (see Appendix A), that $m^R(\mathbf{\Omega})\simeq m(\mathbf{\Omega})$ and $m^I(\mathbf{\Omega})\simeq 0$\footnote{We note that $m^R(\mathbf{\Omega})\simeq m(\mathbf{\Omega})$ and $m^I(\mathbf{\Omega})\simeq 0$ implies that, if one expresses the multiplicative biases as $m_1(\mathbf{\Omega})\gamma_1(\mathbf{\Omega})+{\rm i}m_2(\mathbf{\Omega})\gamma_2(\mathbf{\Omega})$ (where $1$ and $2$ denote the ellipticity components measured parallel to Cartesian axes in a measurement frame, and measured at $45$ degrees to these axes),   $m(\mathbf{\Omega})=m_1(\mathbf{\Omega})=m_2(\mathbf{\Omega})$. This is found to be the case in state-of-the-art methods, e.g. \cite{pujol}. When considering residual systematic effects, after calibration with simulations, this is also expected to be the case.}.} 

Already from the expressions in equation (\ref{shtransform}) it can be seen that multiplicative biases in general mix E and B-modes together, both from the underlying shear field and the multiplicative bias field, and the propagation of such terms is in the form of a convolution represented as a sum over wavenumbers. Furthermore the window function caused by multiplicative biases is $\ell$ and $m$-mode dependent since in general these are not isotropic on the celestial sphere. 

We note that in this case the constant additive biases $c_{1,0}$ and $c_{2,0}$ do not appear in equation (\ref{shtransform}). This is because a constant term only affects the $\ell=0$ mode, but shear is a spin-2 field where the spherical harmonic transform is not defined for $\ell<2$; because 
$_{s}Y_{\ell m}(\mathbf{\Omega})=0$ for $\ell<|s|$. Therefore a constant additive bias cannot affect the cosmic shear power spectrum.

\subsection{Biased Cosmic Shear Power Spectra}
We now compute the expressions for the biased cosmic shear power spectra by taking the correlation of the expressions in equation (\ref{shtransform}). The full expression can be written as a series of terms that pertain to multiplicative, additive and cross-terms, and \extra{depend on the true EE, EB and BB power spectra.} The power spectra estimates are computed by taking the correlation of the spherical harmonic coefficients from equation (\ref{shtransform}) where 
\begin{eqnarray}
\label{eqsize}
\widetilde C^{GH}_{\ell}&\equiv& \frac{1}{2\ell+1}\sum_m 
\widetilde\gamma^G_{\ell m}\widetilde\gamma^{H,*}_{\ell m}
\end{eqnarray}
for $G=(E, B)$ and $H=(E, B)$. 

We provide the full expanded expression for the biased power spectra in Appendix B. If we assume that $C^{EB}_{\ell}=0$, which is the case in all but the most exotic dark energy models, then the three \henk{estimated} power spectra (EE, BB and EB) are:
\begin{eqnarray}
\label{full}
\widetilde C^{EE}_{\ell}&=&
(1+2m_0+m_0^2)C^{EE}_{\ell}\nonumber\\
&+&(1+m_0)({\mathcal N}^+_{\ell}+{\mathcal N}^{+,*}_{\ell})C^{EE}_{\ell}\nonumber\\
&+&
2(1+m_0)C^{c_EE}_{\ell}+
C^{c_Ec_E}_{\ell}\nonumber\\
&+&
\sum_{\ell'}
[{\mathcal M}^{++}_{\ell\ell'}C^{EE}_{\ell'}+
{\mathcal M}^{--}_{\ell\ell'}C^{BB}_{\ell'}]\nonumber\\
&+&\sum_{\ell'}[
{\mathcal B}^{+EE}_{\ell\ell'}+
({\mathcal B}^{+EE}_{\ell\ell'})^*+
{\mathcal B}^{-BE}_{\ell\ell'}+
({\mathcal B}^{-BE}_{\ell\ell'})^*]\nonumber\\
\widetilde C^{BB}_{\ell}&=&
(1+2m_0+m_0^2)C^{BB}_{\ell}\nonumber\\
&+&
(1+m_0)({\mathcal N}^+_{\ell}+{\mathcal N}^{+,*}_{\ell})C^{BB}_{\ell}\nonumber\\
&+&
2(1+m_0)C^{c_BB}_{\ell}+
C^{c_Bc_B}_{\ell}\nonumber\\
&+&
\sum_{\ell'}
[{\mathcal M}^{--}_{\ell\ell'}C^{EE}_{\ell'}+
{\mathcal M}^{++}_{\ell\ell'}C^{BB}_{\ell'}]\nonumber\\
&+&\sum_{\ell'}[
{\mathcal B}^{+BB}_{\ell\ell'}+
({\mathcal B}^{+BB}_{\ell\ell'})^*-
{\mathcal B}^{-EB}_{\ell\ell'}-
({\mathcal B}^{-EB}_{\ell\ell'})^*]\nonumber\\
\widetilde C^{EB}_{\ell}&=&
-(1+m_0){\mathcal N}^{-,*}_{\ell}C^{EE}_{\ell}+
 (1+m_0){\mathcal N}^{-}_{\ell}C^{BB}_{\ell}\nonumber\\
&+&2(1+m_0)C^{c_EB}_{\ell}+C^{c_Ec_B}_{\ell}\nonumber\\
&+&
\sum_{\ell'}
[{\mathcal M}^{-+}_{\ell\ell'}C^{BB}_{\ell'}-
{\mathcal M}^{+-}_{\ell\ell'}C^{EE}_{\ell'}]\nonumber\\
&+&\sum_{\ell'}[
{\mathcal B}^{+EB}_{\ell\ell'}+
({\mathcal B}^{+BE}_{\ell\ell'})^*+
{\mathcal B}^{-BB}_{\ell\ell'}+
({\mathcal B}^{-EE}_{\ell\ell'})^*].\nonumber\\
\end{eqnarray}
The various terms in the full expression are
\begin{eqnarray}
\label{matrices}
    {\mathcal M}^{XY}_{\ell\ell'}&=&\frac{1}{2\ell+1}\sum_{mm'} W^X_{\ell\ell'm m'}(W^Y_{\ell\ell' m m'})^*\nonumber\\
    {\mathcal N}^X_{\ell}&=&\frac{1}{2\ell+1}\sum_{m} W^X_{\ell\ell m m}\nonumber\\
    {\mathcal B}^{XGH}_{\ell\ell}&=&\frac{1}{2\ell+1}\sum_{mm'} W^X_{\ell\ell'm m'}\gamma^G_{\ell'm'} (c^H_{\ell m})^*, 
\end{eqnarray}
where $X=(+,-)$, $Y=(+,-)$, $G=(E, B)$ and $H=(E, B)$. The power spectra in the full expression are labelled in their superscripts as either correlations between shear coefficients ($EE$, $EB$, $BB$), correlations between the additive bias terms ($c_E c_E$,  $c_E c_B$, $c_B c_B$), or cross correlations between shear and additive bias terms ($c_E E$,  $c_E B$, $c_B B$). Equation (\ref{full}) should be defined as the measured power spectrum (on the left hand sides), compared to the power spectrum that would have been measured with no systematic effects (the $C^{GH}_{\ell}$'s on the right hand sides). However we note that the terms convolved with the window function ($W^X_{\ell\ell'm m'}$ in ${\mathcal M}$ and ${\mathcal N}$) in equation (\ref{full}) are derived by taking the ensemble-average of equation (\ref{eqsize}), and making use of the statistical rotational invariance of the ensemble-averaged harmonic modes. Therefore equation (\ref{full}) is a hybrid of ensemble-averaged terms and un-averaged terms which may be non-zero only for a given realisation (as is the case in the examples shown in Section \ref{Simple Simulations}). 

It can be shown that the ${\mathcal N}$ terms are simply the mean of the spatially varying multiplicative bias field. If we consider ${\mathcal N}^+_{\ell}$ we find 
\begin{eqnarray}
    {\mathcal N}^+_{\ell}&=&\frac{1}{(2\ell+1)}
    \sum_m
    \int{\rm d}\mathbf{\Omega}\, m^R(\mathbf{\Omega})_{2}Y^*_{\ell m}(\mathbf{\Omega})_{2}Y_{\ell m}(\mathbf{\Omega})
\end{eqnarray}
We can simplify this expression further by using the generalised addition theorem for spin-weighted spherical harmonics \cite{grain} 
\begin{equation}
\sum_m {}_{s}Y^*_{\ell m}(\mathbf{\Omega}) {}_{s'}Y_{\ell' m'}(\mathbf{\Omega}')=\left[\frac{(2\ell+1)}{4\pi}\right](-1)^{s-s'} D^{\ell}_{ss'}(\alpha, \beta,\gamma)\,{\rm e}^{-2{\rm i}s\gamma},
\end{equation}
where $(\alpha,\beta,\gamma)$ are Euler angles between $\mathbf{\Omega}$ and $\mathbf{\Omega}'$ which in our case are zero, and $D^{\ell}_{ss'}(\alpha, \beta,\gamma)$ are the Wigner rotation matrices which for $D^{\ell}(0,0,0) = \delta^K_{ss′}$. This leads to
\begin{eqnarray}
\label{elin}
{\mathcal N}^+_{\ell}=
\frac{1}{4\pi}\int {\rm d}\mathbf{\Omega}\, m^R(\mathbf{\Omega})=\langle m^R(\mathbf{\Omega})\rangle,
\end{eqnarray}
and similarly ${\mathcal N}^-_{\ell}=\langle m^I(\mathbf{\Omega})\rangle$.
We choose to keep $m_0$ and the mean of $m(\mathbf{\Omega})$ separate since these could have different physical origins i.e. one is a true constant, the other the mean of a spatially varying field. We note that the sum of $m_0$ and $\langle m(\mathbf{\Omega})\rangle$ is similar to the bias $b_m$ term in \cite{tk}.

We discuss further simplifications of these expressions below. In Appendix~C we show the generalisation of this to the case of multiple tomographic bins. 
\begin{figure*}
\centering
{\bf case 1}\quad\quad\quad\quad\quad\quad\quad\quad\quad\quad\quad\quad\quad\quad\;\;\;{\bf case 2}\;\;\;\quad\quad\quad\quad\quad\quad\quad\quad\quad\quad\quad\quad\quad\quad{\bf case 3}\\
\includegraphics[width=0.64\columnwidth]{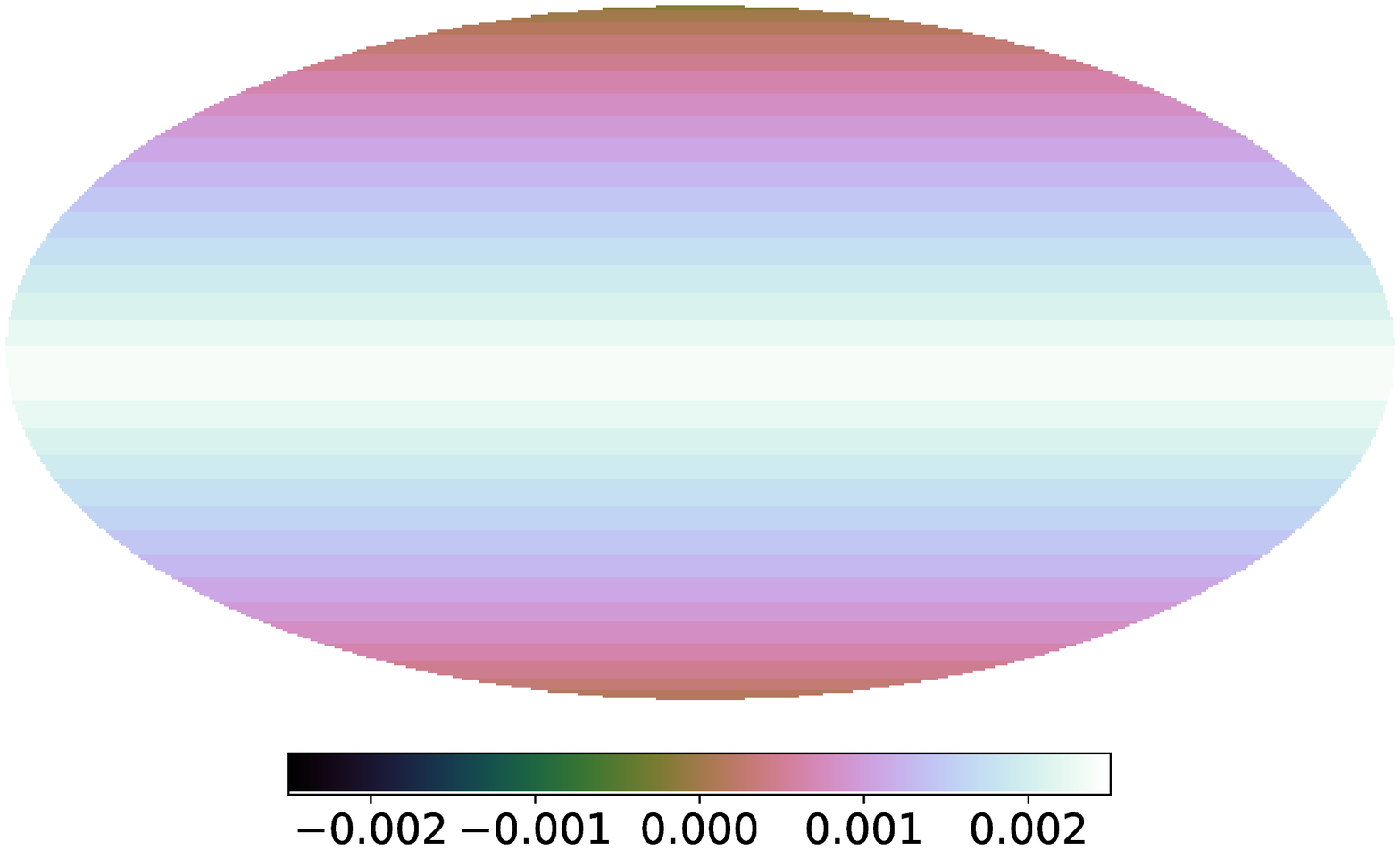}
\includegraphics[width=0.64\columnwidth]{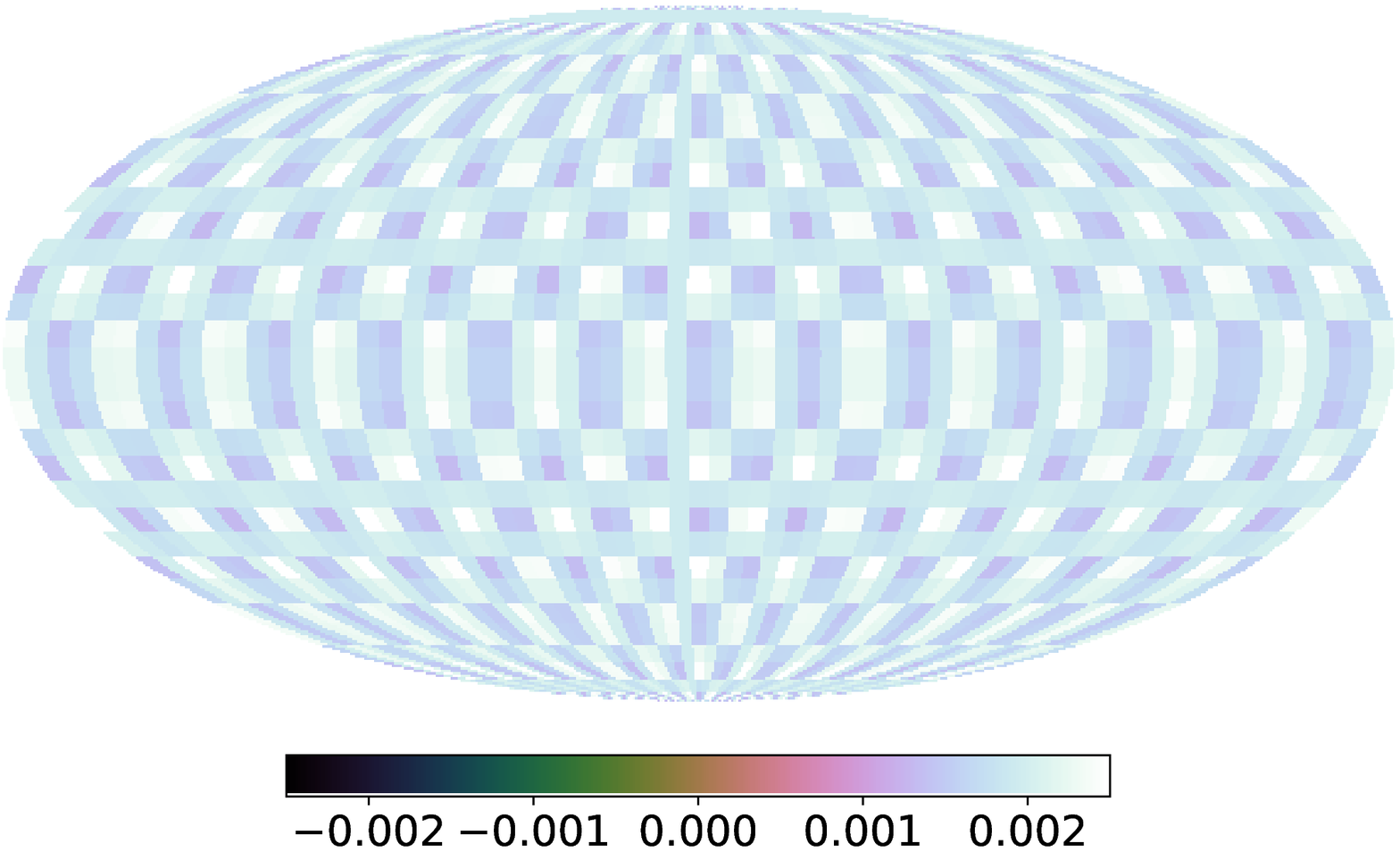}
\includegraphics[width=0.64\columnwidth]{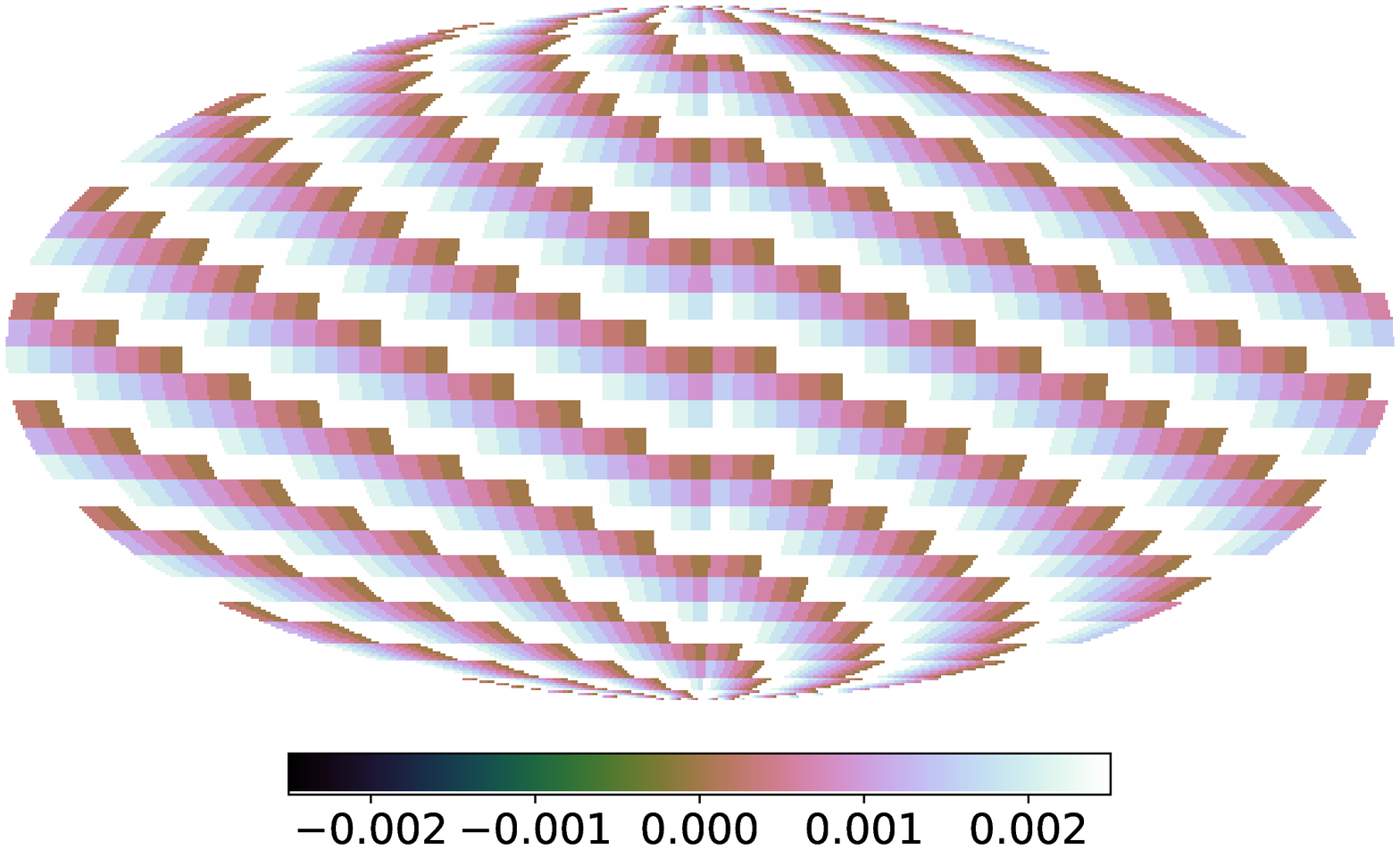}
\caption{The real part of the multiplicative field $m^R(\mathbf{\Omega})$, in the three example cases investigated. Shown is a simulated celestial sphere in a Mollweide Projection with $\theta=\phi=0$ at the North pole. The colour scale represents the amplitude of the biases.}
\label{fig:fields}
\end{figure*} 

\subsubsection{Discussion of the terms}
\label{Discussion of the terms}
We can now discuss each term in the full expression and its physical meaning. 
\begin{itemize}
    \item $m_0$ and $m_0^2$: These terms are the normal contribution from the constant multiplicative bias terms.
    \henk{These arise from the limitations with which shape measurement algorithms can be calibrated \citep[see e.g.][]{Hoekstra17}. As we are concerned with the residual biases \emph{after} such a calibration, $m_0^2\ll m_0$ as $m_0\lesssim 2\times 10^{-3}$ \citep{cropper}}
    \item ${\bf \mathcal N}$: These terms represent multiplicative biases, and are the mean of the multiplicative bias field.   
    \item ${\bf \mathcal M}$: These terms represent multiplicative terms of order $m^2$. The rows and columns show how the E and B-mode power have mixed terms, where the $++$ terms pick up contributions from the real part of the multiplicative bias field, the $--$ terms pick up contributions from the imaginary part, and $+-$ or $-+$ are mixed terms. 
    \item ${\bf \mathcal B}$: These terms represent third-order, bispectrum-like, correlations between the position-dependent multiplicative bias $m(\mathbf{\Omega})\gamma(\mathbf{\Omega})$ and the position-dependent additive bias $c(\mathbf{\Omega})$. Such effects are likely since areas in a survey, or particular pointings, that have detector, telescope or background effects that \henk{cause additive biases will also lead to multiplicative biases. This is because any anisotropic change in the quadrupole moments will modify the size, and thus the multiplicative bias.} We note that note that in this term the multiplicative and shear terms are always spatially coupled and this combination is correlated with the spatially varying additive term.
    \item $c_E E$, $c_E B$, and $c_B B$: these terms capture the correlations between the underlying shear field and the additive bias terms. Such terms are expected to be caused by selection effects in a real survey, where for example blending in high shear regions (e.g. about clusters) could cause an additive bias contribution. 
    \item $c_E c_E$, $c_E c_B$, and $c_B c_B$: These are the power spectra of the position-dependent additive biases. We note again that constant additive bias terms do not contribute to cosmic shear power spectra. 
\end{itemize}

\subsection{Linear Expressions}
\label{Linearised Terms}
Here we show the linearised expressions of the biased cosmic shear power, that only include terms that are linear in the bias parameters. We find that 
\begin{eqnarray}
\label{linear}
\widetilde C^{EE}_{\ell}
&\approx&
(1+2m_0)C^{EE}_{\ell}+2\langle m^R(\mathbf{\Omega})\rangle C^{EE}_{\ell}+2C^{c_EE}_{\ell},\nonumber\\
\widetilde C^{EB}_{\ell}
&\approx&
-\langle m^I(\mathbf{\Omega})\rangle C^{EE}_{\ell}+\langle m^I(\mathbf{\Omega})\rangle C^{BB}_{\ell}+
2C^{c_EB}_{\ell},\nonumber\\
\widetilde C^{BB}_{\ell}
&\approx&
(1+2m_0)C^{BB}_{\ell}+2\langle m^R(\mathbf{\Omega})\rangle C^{BB}_{\ell}+2C^{c_BB}_{\ell}.
\end{eqnarray}
We have included $B$-mode power since as shown in \cite{Schneider02} source redshift clustering can cause a small $B$-mode component. 

We see that the impact of spatially varying biases will be, to linear order, captured by the mean of the multiplicative bias and the additive-shear cross correlation power spectrum, \henk{but in the presence of intrinsic B-modes, $\widetilde C^{EB}_{\ell}$ now includes a term $2\langle m^I(\mathbf{\Omega})\rangle C^{BB}_{\ell}$.} 

We note that if $\langle m^R(\mathbf{\Omega})\rangle=\langle m^I(\mathbf{\Omega})\rangle$, then (twice) the EB power spectrum could be added to the EE power spectrum to cancel out multiplicative effects; however this is not expected to be the case in general, or for small biases.

\section{Simple Simulations}
\label{Simple Simulations}
To test that the above formalism can indeed capture the propagation of general position-dependent multiplicative and additive bias terms into the cosmic shear power spectrum we generate several toy examples and investigate the contributions of each term to the overall change. For each case we define a multiplicative constant and field, $m_0$ and $m(\mathbf{\Omega})$, and an additive constant and field, $c_0=c_{1,0}=c_{2,0}$ and $c(\mathbf{\Omega})$, although the choice for $c_0$ has no impact on cosmic shear power spectra by definition. We normalise these fields such that 
$\langle m_0 + m(\mathbf{\Omega})\rangle = 2\times 10^{-3}$ and $\langle c_0 + c(\mathbf{\Omega})\rangle = 1\times 10^{-4}$, which represent the overall requirements for a \emph{Euclid}-like experiment \citep{cropper}; however we note that the amplitude of $\langle c_0 + c(\mathbf{\Omega})\rangle$ will have no effect on the power spectrum as discussed previously. For each case we compare the computation of the analytic expression in equation (\ref{full}) and a numerical case where we compute the real space shear field $\widetilde \gamma(\mathbf{\Omega})=[1+m_0+m(\mathbf{\Omega})]\gamma(\mathbf{\Omega})+[c_0+c(\mathbf{\Omega})]$ and then compute the power spectra of this directly using a spherical harmonic transform. In all cases we compute the original $\gamma(\mathbf{\Omega})$ field using a Gaussian random field generated by using a cosmic shear power spectrum based on the {\it Planck} $\Lambda$CDM cosmology \citep{planck}, using the {\tt massmappy} code \citep{wallis}. In all cases we use {\tt SSHT} \cite{ssht} to compute the spin-weighted spherical harmonics, which sample the sphere using the sampling scheme of \cite{mw}. 

The cases we consider are shown below. Note that we express these in terms of an arbitrary amplitude $A$ since these are all normalised to have $\langle m_0 + m(\mathbf{\Omega})\rangle = 2\times 10^{-3}$ and $\langle c_0 + c(\mathbf{\Omega})\rangle = 1\times 10^{-4}$. The cases are simple examples but nonetheless are approximations of realistic spatial variations that could occur: 
\begin{enumerate}
    \item Case 1, Simple Galactic Plane: 
    \begin{itemize}
     \item $m^R(\mathbf{\Omega})=A[\pi-|\phi-\pi|]$, $m^I(\mathbf{\Omega})=0$, 
     \item $c^R(\mathbf{\Omega})=c^I(\mathbf{\Omega})=A[\pi-|\phi-\pi|]$,
     \item $c_0=A$, $m_0=A$; 
   \end{itemize}   
    \item Case 2, Simple Patch Pattern:
    \begin{itemize}
    \item $m^R(\mathbf{\Omega})=10A\sin(100|\phi-\pi|)\sin(100|\theta-\pi|)$, $m^I(\mathbf{\Omega})=0$, 
    \item $c^R(\mathbf{\Omega})=c^I(\mathbf{\Omega})=10A\sin(10|\phi-\pi|)\sin(10|\theta-\pi|)$, 
    \item $c_0=A$, $m_0=A$; 
    \end{itemize}
    \item Case 3, Simple Scanning Pattern:
    \begin{itemize}
    \item $m^R(\mathbf{\Omega})=Ai$, where i is an iterative pixel number count, which is reset when $i=10$, $m^I(\mathbf{\Omega})=0$, 
    \item $c^R(\mathbf{\Omega})=c^I(\mathbf{\Omega})=Ai^2$, where i is an iterative pixel number count, which is reset when $i=10$,
    \item 
    $c_0=A$, $m_0=A$.
    \end{itemize}
\end{enumerate}
\begin{figure*}
\centering
\includegraphics[width=0.714\columnwidth]{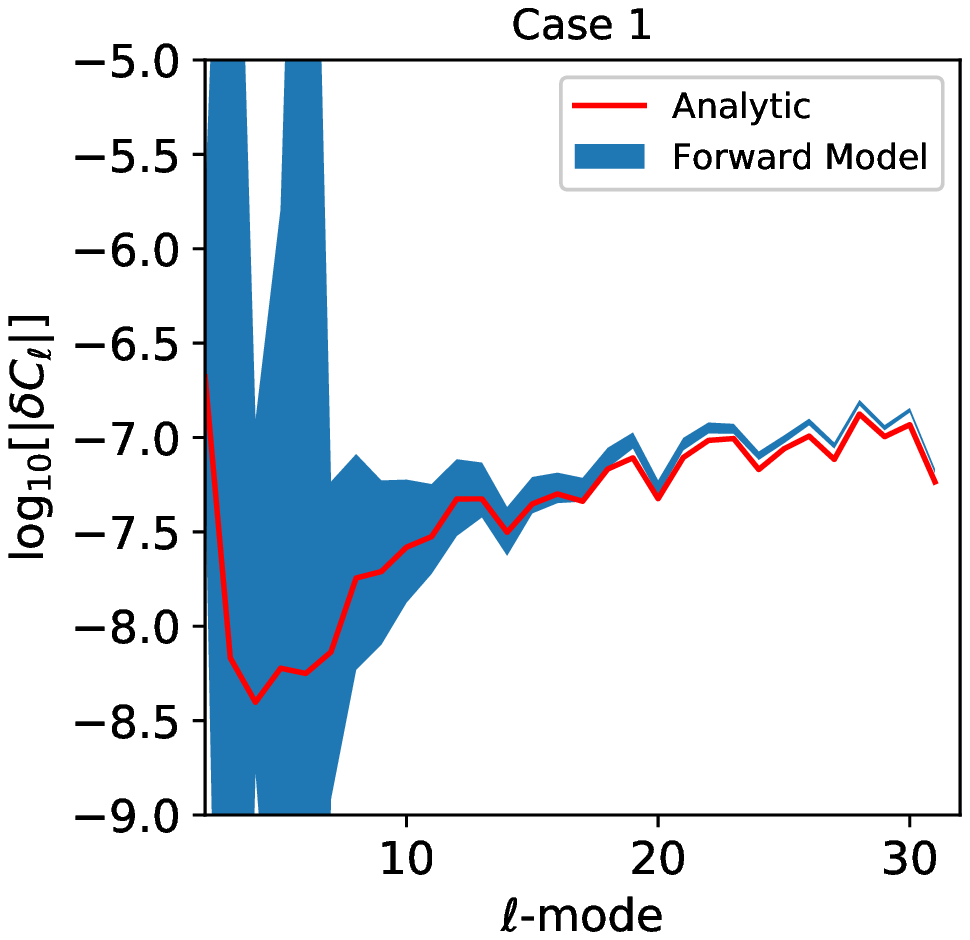}
\includegraphics[width=\columnwidth]{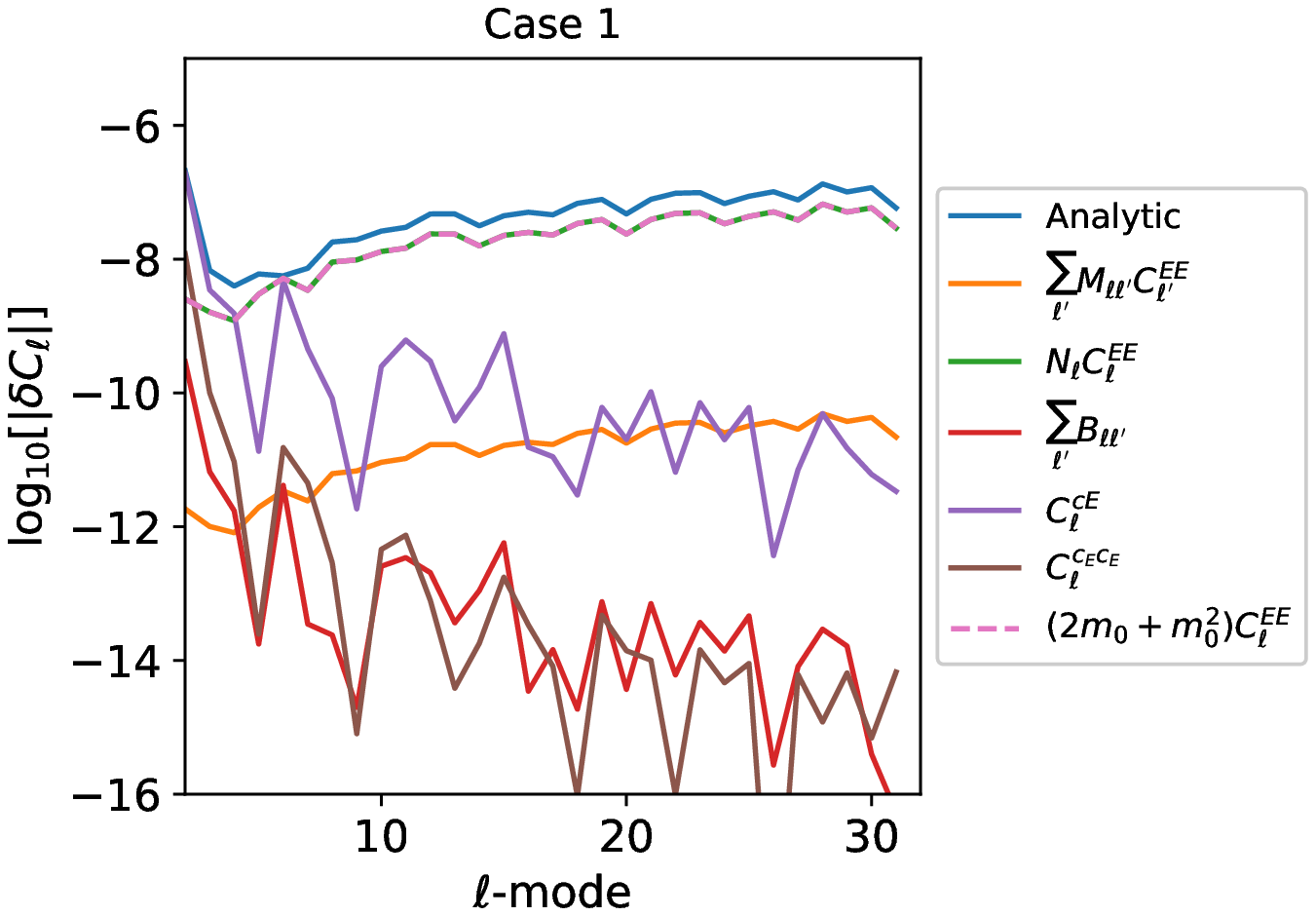}\\
\includegraphics[width=0.714\columnwidth]{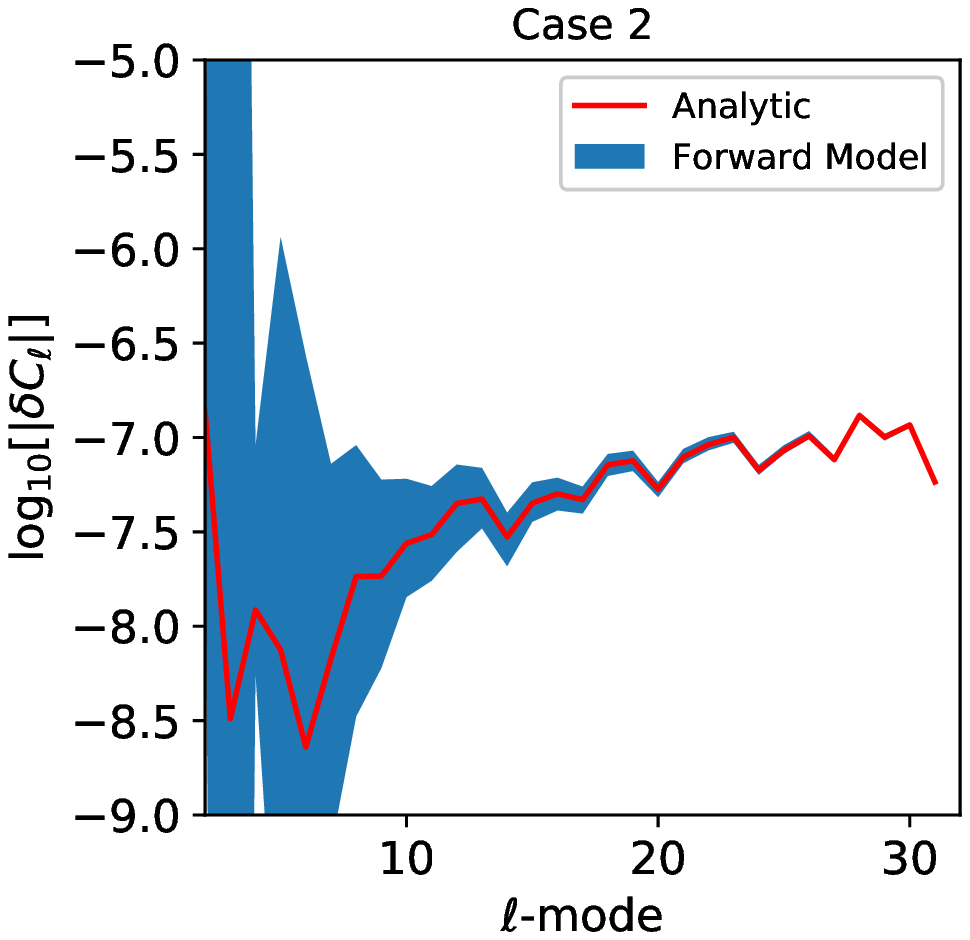}
\includegraphics[width=\columnwidth]{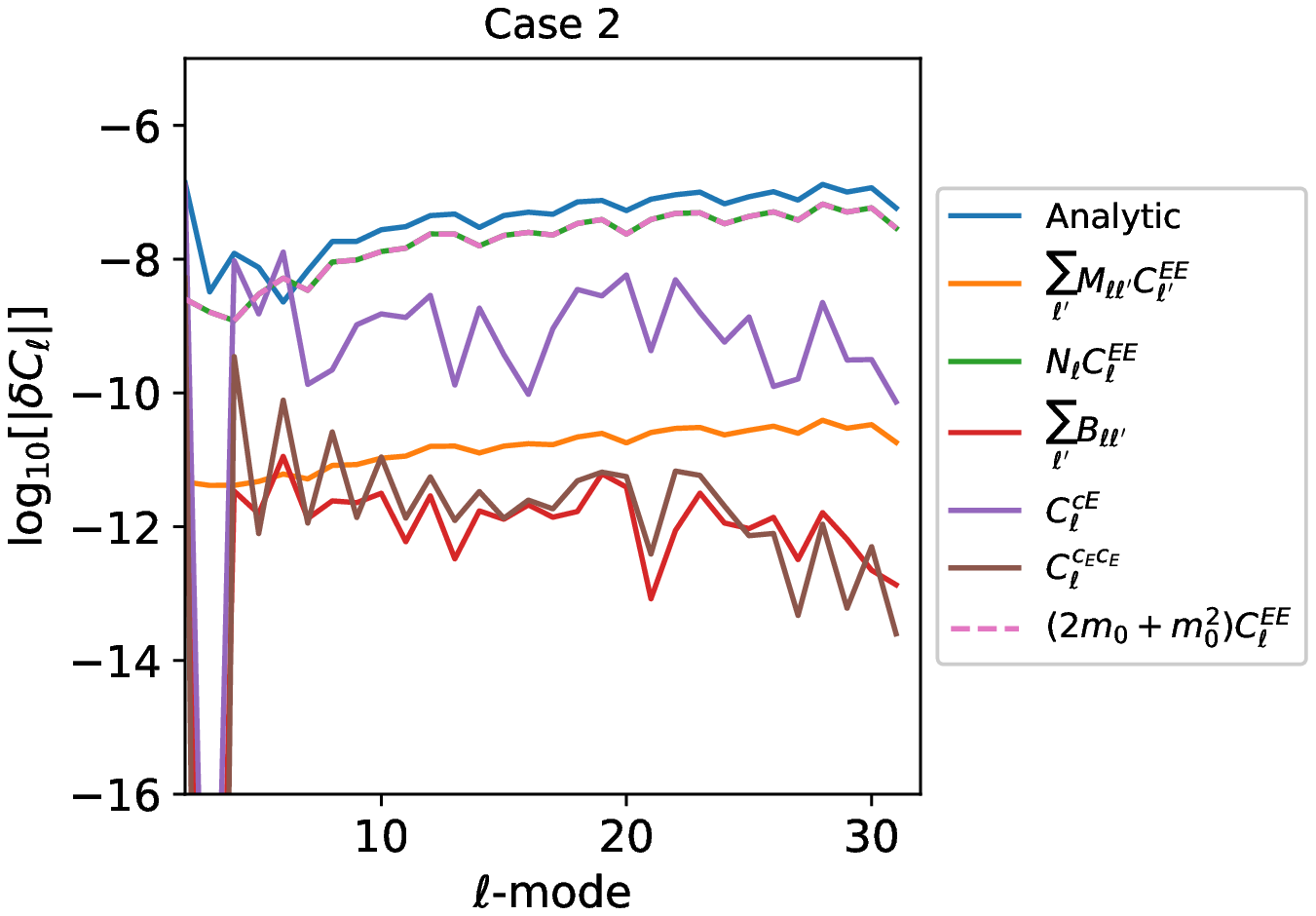}\\
\includegraphics[width=0.714\columnwidth]{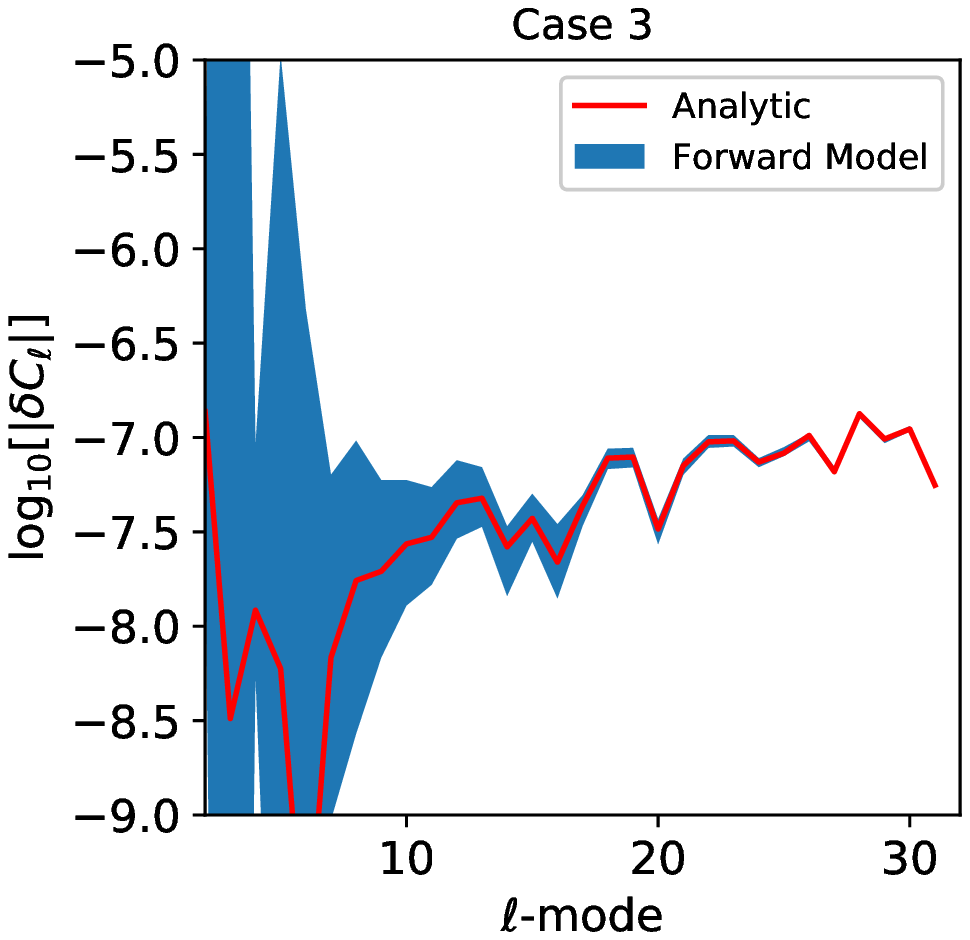}
\includegraphics[width=\columnwidth]{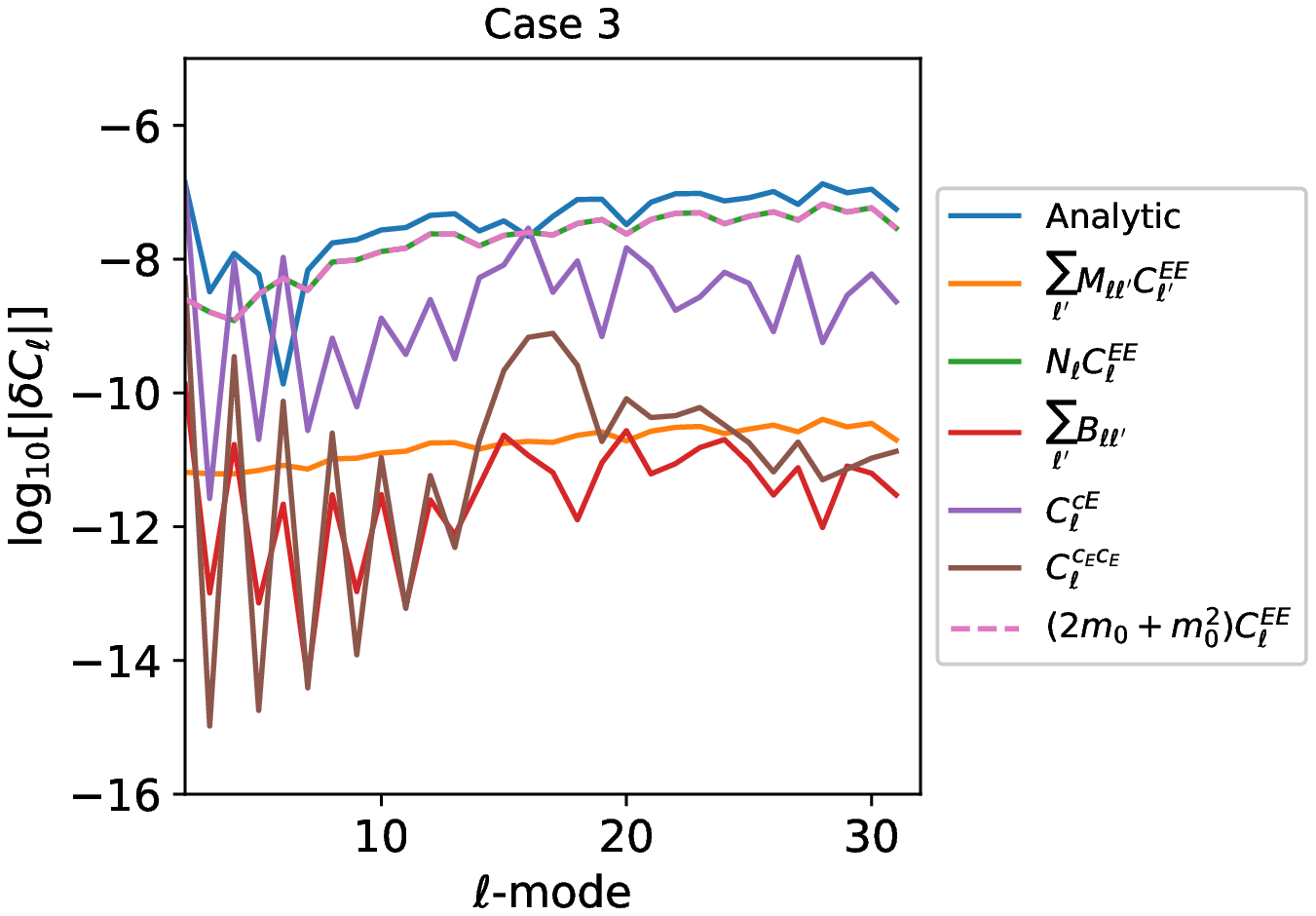}
\caption{The residual power spectrum $\delta C_{\ell}=\widetilde C^{EE}_{\ell}-C^{EE}_{\ell}$ for the three cases considered. The left plots show the comparison between the numerical case computed by transforming the modified shear field and performing a spherical harmonic transform (i.e. a forward model), and the analytic case computed using equation (\ref{full}). The blue band shows the $1$-sigma scatter about the mean of forward model, the red lines show the analytic prediction. The right plots show the contribution to the analytic case from each of the components in equation (\ref{full}). In all cases the legends label the coloured lines. The mean multiplicative terms (green and pink lines) have the same value due to the scaling, and hence are over-plotted.}
\label{fig:power}
\end{figure*}

The first case approximates a Galactic plane dependency, the second case approximates a patch-dependent systematic effect, and the third case is a non-analytic case that approximates a scanning sequence of exposures. To demonstrate the complexity of the spatial variation of the cases we show in Figure \ref{fig:fields} the real part of the multiplicative field for each of the cases (we do not show all the fields associated with the systematic effects since they are largely similar in form).

In Figure \ref{fig:power} we show the residual power spectra 
$\delta C_{\ell}=\widetilde C^{EE}_{\ell}-C^{EE}_{\ell}$ for each of the cases considered. We compute the error on the forward model power spectrum as $\sigma(\delta C_{\ell})=[(\widetilde C^{EE}_{\ell})^2+(C^{EE}_{\ell})^2]^{1/2}$ \citep{jb, hj}.

We note that in all cases we use a maximum angular multipole of $L=32$. This is because the calculations are particularly numerically demanding. The $W^{\pm}_{\ell\ell'm m'}$ calculations have dimension $L^4$, and  for each of these spin-weighted spherical harmonic functions must be computed each of which scale like $L^2\log L$ at best \citep{ssht}. This point is discussed further in Section \ref{s:scaling}. In all cases the analytic formula given in equation (\ref{full}) accurately captured the form of the residual power spectrum; the very small differences are due to the numerical stability of the spin-weighted spherical harmonic transform calculations. 
 
\begin{figure}
\centering
\includegraphics[width=\columnwidth]{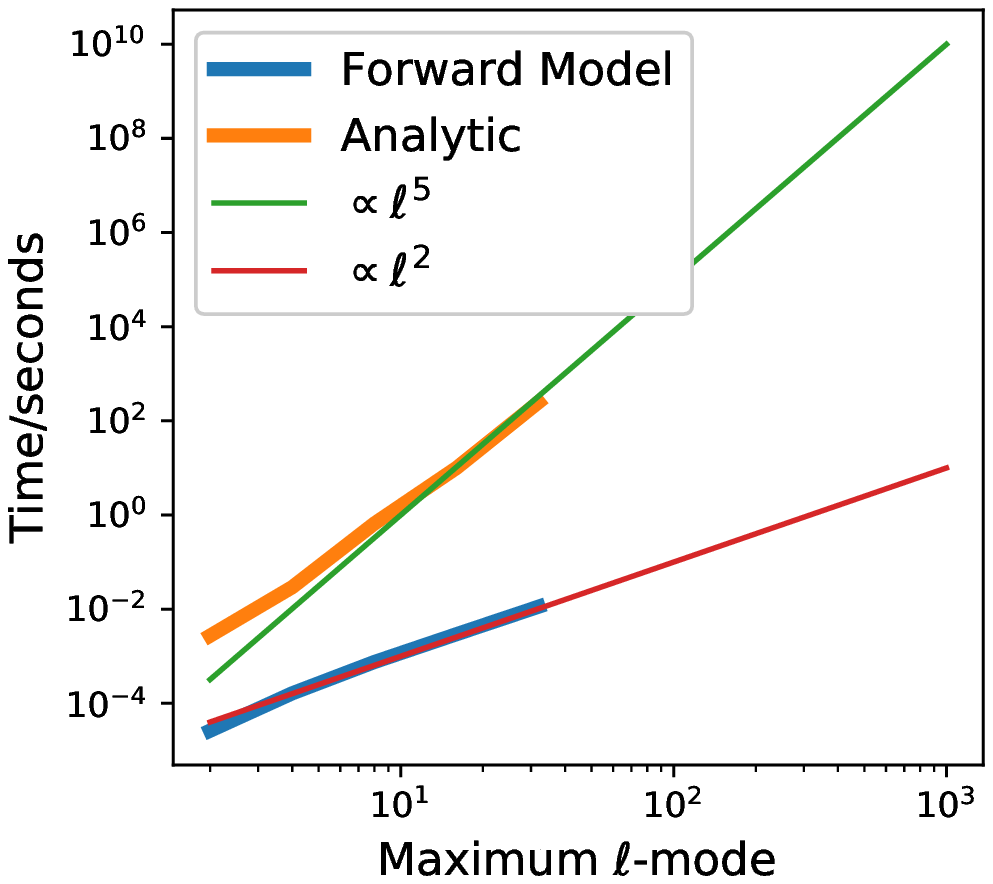}
\caption{Timing of the forward model and analytic calculations
using a 2016 Macbook Pro, 3.3 GHz Intel Core i7, 16 GB 2133 MHz LPDDR3. The blue line shows the forward model scaling, the orange line shows the analytic scaling, the thin red and green lines are proportional to $L^2$ and $L^5$ respectively where $L$ is the maximum $\ell$-mode.}
\label{f:scaling}
\end{figure} 

With regard to the different terms we find in all cases that the ${\mathcal N}C^{EE}_{\ell}$ term is dominant, which is expected since it is of linear order in $m(\mathbf{\Omega})$, followed by the $C^{c_EE}_{\ell}$ cross-correlation term and the $(m_0+m_0^2)C^{EE}_{\ell}$ terms. The convolutive ${\mathcal M}$, ${\mathcal B}$ and $C^{c_E c_E}_{\ell}$ terms are all at least an order of magnitude lower in all cases. Therefore the linearised expression in equation (\ref{elin}) $\delta C_{\ell}\approx
2m_0C^{EE}_{\ell}+2\langle m^R(\mathbf{\Omega})\rangle C^{EE}_{\ell}+2C^{c_EE}_{\ell}$ is a good approximation in these simple examples. In the case that the mean of $m(\mathbf{\Omega})$ and $c(\mathbf{\Omega})$ are both zero, all of the remaining terms at second and third order in bias would become important at approximately the same level. Note that we plot the absolute value of the residual power spectrum contributions, since some terms can be negative depending on the nature of spatial pattern used in the simulations.

\section{Discussion}
\label{Discussion}
We have shown in general how constant and position-dependent shape measurement biases propagate through to cosmic shear power spectra. 

The multiplicative bias terms shown here are similar to those that result in CMB polarisation pseudo-$C_{\ell}$ analyses, where masking of the data results in expressions that also include temperature power spectra \citep{lct, zs, grain, bct}. The difference here is that instead of a mask we have a multiplicative bias field that is in general spin-dependent. We note  that this formalism equally applies to the case of masked cosmic shear data where $m(\mathbf{\Omega})$ may be zero in some regions, and that such a case would lead to further mode mixing. 

In \cite{great10} a pseudo-$C_{\ell}$ formalism was used to assess position-dependent shape measurement errors. However in that study the linear  terms ${\mathcal N}$, bispectrum and additive terms were not included and the non-linear convolution term ${\mathcal M}$ used a simpler form based on a flat-sky approximation. In \cite{tk} the propagation of shape measurement biases was generalised to the convolutive case but the linear terms, bispectrum and E/B mode mixing terms were ignored. 

In \cite{massey} and \cite{cropper} requirements were set on weak lensing experiments using an approximation for position-dependent shape measurement biases. In that case a form of propagation was determined for the constant case $\widetilde C^{EE}_{\ell}=(1+2m_0+m_0^2)C^{EE}_{\ell}+c^2_c$, which was then replaced with a `position-dependent' formulation proposed by \cite{AR},
$C^{EE}_{\ell}=(1+{\mathcal M}_{\ell})C^{EE}_{\ell}+{\mathcal A}_{\ell}$. We find that such an expression is similar to the full case when only linear terms in biases are assumed, whereas the relationship to the underlying position-dependent bias fields is much more complex. Furthermore in \cite{massey} a worst-case scenario in sensitivity was assumed for ${\mathcal M}_{\ell}$ in which multiplicative biases \henk{mimicked the scale-dependent behaviour of dark energy.} These worst-case assumptions are conservative when designing an experiment and lead to requirements that will guarantee performance, but when assessing the actual performance of a survey they are not adequate. 

We note that if the mean multiplicative bias is zero -- as may be expected if pre-experiment simulations can determine any mean effect -- then only second and third order convolutive terms remain. The power spectrum residuals caused by these terms, and the impact on cosmology, are expected to be much lower than than the mean terms for two reasons. First, because the terms are second order and so for $m(\mathbf{\Omega})\ll 1$ these are smaller. Second, because they are convolutions it is unlikely that a functional form will result that matches the cosmic shear power spectrum; therefore the impact on cosmological parameter inference is expected to be lower \citep{tk}.

\newpage
\subsection{Scaling}
\label{s:scaling}
In Figure \ref{f:scaling} we show how the analytic and forward modelling cases scale as a function of $L$, the maximum $\ell$-mode . To compute the full case requires evaluations of terms that scale like $L^4\times L^2\log L$, where $L$ is the maximum multipole; this is because of the $W_{\ell\ell' m m'}$ terms that have $\sim L^4$ summations, and the spherical harmonic transforms that scale like $L^2\log L$. We find a slightly better scaling due to pre-computation of the spin-weighted spherical harmonic functions, but nonetheless the analytic calculation scales like $\propto L^5$ compared to the forward modelling that scales like $\propto L^2$.  

We evaluate simple examples for $L=32$, but scaling to a reasonable value of $L>1000$ would result in prohibitively long calculations. On the other hand the forward modelling of systematic effects i.e. the evaluation of equation (\ref{gamma}) in real space and a direct spherical harmonic transform to produce a power spectrum is tractable for $L>1000$ and we therefore advocate this approach in \cite{2019arXiv190405364T} and Paykari et al., (in prep).  

We note that one could perform a spherical sky analysis and supplement this by a fast Fourier transform on small scales. However this approach would require the sphere to be divided into patches upon which a flat sky could be run with overlap between patches to capture all angular modes, and a transition from all-sky to flat-sky computed. This is feasible, but this is unneeded complexity given that a forward model is very simple to compute.

\section{Conclusions}
\label{Conclusions}
In this paper we derive a complete expression for the impact of constant and spatially varying multiplicative and additive shape measurement biases on the cosmic shear power spectrum. In doing so we find several terms that have thus far been overlooked, in particular terms relating to cross-correlations between biases and shear. 

In performing the full calculation we find that to linear order spatially varying biases are well approximated by \henk{the sum of the product of the the power spectrum and the mean of the multiplicative bias field, and the cross-correlation term between the additive bias field and the shear field}. We note that the cosmic shear power spectrum is not sensitive to constant or mean additive biases. 

We compare the computation of the full analytic expression with that obtained using a forward modelling approach using simplified simulations and find good agreement. Furthermore we use these simplified simulations to demonstrate how each term in the full expression contributes to the total. However, in performing the full calculation we also find that its computation scales as the maximum multipole $\propto L^5$. This means that its evaluation for large $L>1000$ is unfeasible. Therefore we recommend that any assessment of the impact of biases on cosmic shear power spectrum must be performed using a forward modelling approach.

\acknowledgements
We thank the developers of {\tt SSHT} and {\tt massmappy}, Jason McEwen, Chris Wallis and Boris Leistedt for making their code publicly available. We thank Dipak Munshi for useful discussions. \peter{We thank Peter Schneider for comments on an early draft.} PP is supported by the UK Science and Technology Facilities Council. TK is supported by a Royal Society University Research Fellowship. HH acknowledges support from Vici grant 639.043.512 financed by the Netherlands Organization for Scientific Research (NWO). We thank an anonymous referee whose comments improved this manuscript.

\bibliographystyle{mnras}
\bibliography{sample.bib}

\appendix

\section{Bias Propagation}
The true shear field $\gamma(\mathbf{\Omega})$ is a spin-2 quantity, and any impact of imperfect shape measurement should preserve this spin-2 nature. In general we consider that there are two ways that an imperfect shape measurement can impact an observed spin-2 field either 1) there can be a incorrect estimate of the ratio of the semi-major axes, expressed as an amplitude change of the shear, 2) there can be an incorrect estimation of the observed angle of the ellipse i.e. a rotation. 
\\
\\
\subsection{Multiplicative Bias}
\noindent For a multiplicative systematic effect these possibilities can be expressed as 
\begin{eqnarray}
[1+m(\mathbf{\Omega})\gamma(\mathbf{\Omega})]=[1+m(\mathbf{\Omega}){\rm e}^{{\rm i}\phi_m(\mathbf{\Omega})}]\,|\gamma(\mathbf{\Omega})|{\rm e}^{{\rm i}2\Phi(\mathbf{\Omega})}
\end{eqnarray}
where we have expressed $\gamma(\mathbf{\Omega})=|\gamma(\mathbf{\Omega})|{\rm e}^{{\rm i}2\Phi(\mathbf{\Omega})}$ where $\Phi(\mathbf{\Omega})$ is the angle between the orientation of the elliptical shape induced by the shear and the $x$-axis of local Cartesian frame in which measurement has been made. To this expression we apply an amplitude change $m(\mathbf{\Omega})$ corresponding to an incorrect measurement of the ratio of the semi-major axes, and a small rotation $\phi_m(\mathbf{\Omega})$. This preserves the spin-2 nature of the measured field. When we express the product of two complex numbers $m(\mathbf{\Omega})\gamma(\mathbf{\Omega})$, it should be understood that the multiplicative bias fields take the form 
\begin{eqnarray}
 m(\mathbf{\Omega})&=&m^R(\mathbf{\Omega})+{\rm i}m^I(\mathbf{\Omega})\nonumber\\
 m^R(\mathbf{\Omega})&=&m(\mathbf{\Omega})\cos[\phi_m(\mathbf{\Omega})]\nonumber\\
 m^I(\mathbf{\Omega})&=&m(\mathbf{\Omega})\sin[\phi_m(\mathbf{\Omega})],
\end{eqnarray}
where $m^R(\mathbf{\Omega})$ and $m^I(\mathbf{\Omega})$ in the first equation are the real and imaginary parts respectively which are coupled as expressed in the subsequent equations. We note that we do not label these $m_1(\mathbf{\Omega})$ and $m_2(\mathbf{\Omega})$ since they do not map solely to the $\gamma_1(\mathbf{\Omega})$ and $\gamma_2(\mathbf{\Omega})$ components of the shear field where $\gamma(\mathbf{\Omega})=\gamma_1(\mathbf{\Omega})+{\rm i}\gamma_2(\mathbf{\Omega})$. 

We refer to \cite{pujol,metacal} for further discussion of the propagation of more complex multiplicative biases. We note that if the amplitude of the (residual) biases are small, and that the rotation angle is random, then it is reasonable to assume that $\langle m^R(\mathbf{\Omega})\rangle=\langle m^I(\mathbf{\Omega})\rangle$; if one assumes instead small residual biases and applies the small angle approximation then $m^R(\mathbf{\Omega})\approx m(\mathbf{\Omega})$ and  $m^I(\mathbf{\Omega})\approx m(\mathbf{\Omega})\phi_m(\mathbf{\Omega})\approx 0$. In the case that $m^R(\mathbf{\Omega})=m(\mathbf{\Omega})$ and $m^I(\mathbf{\Omega})=0$ then this would result in $m_1(\mathbf{\Omega})=m(\mathbf{\Omega})$ and $m_1(\mathbf{\Omega})=m_2(\mathbf{\Omega})$ if expressed in this way. 

We note that if one applies a bias of the form $m_1(\mathbf{\Omega})\gamma_1(\mathbf{\Omega})+{\rm i}m_2(\mathbf{\Omega})\gamma_2(\mathbf{\Omega})$ (i.e. a different independent scalar multiplicative biases applied to each of the shear components) then this cannot in general be expressed as an amplitude change with a rotation, and therefore can result in a change in the spin properties of the measured field. We note that one could create such an expression by $m'(\mathbf{\Omega})\gamma(\mathbf{\Omega})+\delta m(\mathbf{\Omega})\gamma^*(\mathbf{\Omega})$ where, $m'(\mathbf{\Omega})=[m_1(\mathbf{\Omega})+m_2\mathbf{\Omega})]/2$ and $\delta m(\mathbf{\Omega})=[m_1(\mathbf{\Omega})-m_2(\mathbf{\Omega})]/2$, however the second term would represent a parity change/mislabelling in the $\gamma_2$ component which would be a very large systematic effect. We explored the expected size of $\delta m/m$ using image simulations that resemble \emph{Euclid} based on \cite{Hoekstra17} and found that $\delta m/m\sim 0.1$ for both PSF anisotropy and for a simple of charge trailing between pixels. Hence in practice it appears from these initial studies that on can typically ignore $\delta m$. 
\\
\\
\subsection{Additive Bias}
\noindent In the additive bias case one can add a field with spin-2 properties such that 
\begin{eqnarray}
    \gamma(\mathbf{\Omega})+c(\mathbf{\Omega})=
    \gamma(\mathbf{\Omega})+|c(\mathbf{\Omega})|{\rm e}^{{\rm i}\phi_c(\mathbf{\Omega})}{\rm e}^{{\rm i}2\Phi(\mathbf{\Omega})}
\end{eqnarray}
where $|c(\mathbf{\Omega})|$ and $\phi_c(\mathbf{\Omega})$ are systematic changes in the amplitude and rotation angle of the measurements respectively. In this case the two additive components can be expressed like 
\begin{eqnarray}
    c(\mathbf{\Omega})&=&c_1(\mathbf{\Omega})+{\rm i}c_2(\mathbf{\Omega})\nonumber\\
    c_1(\mathbf{\Omega})&=&|c(\mathbf{\Omega})||\cos(2\Phi(\mathbf{\Omega})+\phi_c(\mathbf{\Omega}))\nonumber\\
    c_2(\mathbf{\Omega})&=&|c(\mathbf{\Omega})|\sin(2\Phi(\mathbf{\Omega})+\phi_c(\mathbf{\Omega})),
\end{eqnarray}
where in the additive case the real and imaginary parts will add to the respective  $\gamma_1$ and $\gamma_2$ parts of the shear and hence we label them as such (which is not the case for the multiplicative biases). In the constant case one can write $c_0=c_{1,0}+{\rm i}c_{2,0}$ where $c_{1,0}$ and $c_{2,0}$ are constants. 

\section{The Two Dimensional Case}
The full expanded expression for the two dimensional case can be written in matrix form as
\begin{eqnarray}
\label{full2}
    \left( \begin{array}{c}
\widetilde C^{EE}_{\ell}\\
\widetilde C^{EB}_{\ell}\\
\widetilde C^{BB}_{\ell}\end{array} \right)
&=& 
(1+2m_0+m_0^2)\left( \begin{array}{c}
C^{EE}_{\ell}\\
C^{EB}_{\ell}\\
C^{BB}_{\ell}\end{array} \right)\nonumber\\
&+&
(1+m_0)
\left( \begin{array}{ccc}
 ({\mathcal N}^+_{\ell}+{\mathcal N}^{+,*}_{\ell})&{\mathcal N}^-_{\ell}&0\\
-{\mathcal N}^{-,*}_{\ell}&({\mathcal N}^+_{\ell}+{\mathcal N}^{+,*}_{\ell})&{\mathcal N}^-_{\ell}\\
0&-{\mathcal N}^{-,*}_{\ell}&({\mathcal N}^+_{\ell}+{\mathcal N}^{+,*}_{\ell})
\end{array} \right)
\left( \begin{array}{c}
C^{EE}_{\ell}\\
C^{EB}_{\ell}\\
C^{BB}_{\ell}\end{array} \right)\nonumber\\
&+&
2(1+m_0)\left( \begin{array}{c}
C^{c_E E}_{\ell}\\
C^{c_E B}_{\ell}\\
C^{c_B B}_{\ell}\end{array} \right)
+
\left( \begin{array}{c}
C^{c_E c_E}_{\ell}\\
C^{c_E c_B}_{\ell}\\
C^{c_B c_B}_{\ell}\end{array} \right)\nonumber\\
&+&
\sum_{\ell'}
\left( \begin{array}{ccc}
 {\mathcal M}^{++}_{\ell\ell'}& ({\mathcal M}^{-+}_{\ell\ell'}+{\mathcal M}^{+-}_{\ell\ell'})&{\mathcal M}^{--}_{\ell\ell'}\\
-{\mathcal M}^{+-}_{\ell\ell'}& ({\mathcal M}^{++}_{\ell\ell'}-{\mathcal M}^{--}_{\ell\ell'})&{\mathcal M}^{-+}_{\ell\ell'}\\
 {\mathcal M}^{--}_{\ell\ell'}&-({\mathcal M}^{-+}_{\ell\ell'}+{\mathcal M}^{+-}_{\ell\ell'})&{\mathcal M}^{++}_{\ell\ell'}\\
\end{array} \right)
\left( \begin{array}{c}
C^{EE}_{\ell'}\\
C^{EB}_{\ell'}\\
C^{BB}_{\ell'}\end{array} \right)\nonumber\\
&+&
\left( \begin{array}{c}
{\mathcal B}^{+EE}_{\ell\ell'}+({\mathcal B}^{+EE}_{\ell\ell'})^*+{\mathcal B}^{-BE}_{\ell\ell'}+({\mathcal B}^{-BE}_{\ell\ell'})^*\\
{\mathcal B}^{+EB}_{\ell\ell'}+({\mathcal B}^{+BE}_{\ell\ell'})^*+{\mathcal B}^{-BB}_{\ell\ell'}+({\mathcal B}^{-EE}_{\ell\ell'})^*\\
{\mathcal B}^{+BB}_{\ell\ell'}+({\mathcal B}^{+BB}_{\ell\ell'})^*-{\mathcal B}^{-EB}_{\ell\ell'}-({\mathcal B}^{-EB}_{\ell\ell'})^*  
\end{array} \right). 
\end{eqnarray}
The ${\mathcal M}$, ${\mathcal N}$ and ${\mathcal B}$ terms are defined in equation (\ref{matrices}) in the main body of the text.

\section{The Tomographic Case}
In equation (\ref{full}) we consider the case of a single population of galaxies, however in reality one typically will define several populations labelled as tomographic bins normally delineated in redshift. In this case equation (\ref{gamma}) is labelled with a tomographic bin $i$ such that 
\begin{eqnarray}
    \widetilde\gamma_i(\mathbf{\Omega})=[1+m_{0,i}+m_i(\mathbf{\Omega})]\gamma_i(\mathbf{\Omega})+[c_{0,i}(1+{\rm i})+c_i(\mathbf{\Omega})]
\end{eqnarray}
and the power spectra are defined as
\begin{eqnarray}
\widetilde C^{EE}_{\ell,ij}&\equiv& \frac{1}{2\ell+1}\sum_m 
\widetilde\gamma^E_{\ell m,i}\widetilde\gamma^{E,*}_{\ell m,j}\nonumber\\
\widetilde C^{BB}_{\ell,ij}&\equiv& \frac{1}{2\ell+1}\sum_m 
\widetilde\gamma^B_{\ell m,i}\widetilde\gamma^{B,*}_{\ell m,j}\nonumber\\
\widetilde C^{EB}_{\ell,ij}&\equiv& \frac{1}{(2\ell+1)}\sum_m 
\widetilde\gamma^E_{\ell m,i}\widetilde\gamma^{B,*}_{\ell m,j}\nonumber\\
\widetilde C^{BE}_{\ell,ij}&\equiv& \frac{1}{(2\ell+1)}\sum_m 
\widetilde\gamma^B_{\ell m,i}\widetilde\gamma^{E,*}_{\ell m,j},
\end{eqnarray}
where $i$ and $j$ label tomographic bins. We use notation where $C^{XY}_{\ell,ij}$ means that field $X$ is associated with $i$, and field $Y$ is associated with $j$. We note that in this case there is a difference between the $EB$ power spectrum and the $BE$ power spectrum for tomographic bins $ij$. The full expression in expanded form is then: 
\begin{eqnarray}
\label{full3}
    \left( \begin{array}{c}
\widetilde C^{EE}_{\ell,ij}\\
\widetilde C^{EB}_{\ell,ij}\\
\widetilde C^{BE}_{\ell,ij}\\
\widetilde C^{BB}_{\ell,ij}\end{array} \right)
&=& 
(1+m_{0,i}+m_{0,j}+m_{0,i}m_{0,j})\left( \begin{array}{c}
C^{EE}_{\ell,ij}\\
C^{EB}_{\ell,ij}\\
C^{BE}_{\ell,ij}\\
C^{BB}_{\ell,ij}\end{array} \right)\nonumber\\
&+&
(1+m_{0,j})
\left( \begin{array}{cccc}
 {\mathcal N}^+_{\ell,i}&0&{\mathcal N}^-_{\ell,i}&0\\
0&-{\mathcal N}^+_{\ell,i}&0&{\mathcal N}^-_{\ell,i}\\
-{\mathcal N}^-_{\ell,i}&0&{\mathcal N}^+_{\ell,i}&0\\
0&-{\mathcal N}^-_{\ell,i}&0&{\mathcal N}^+_{\ell,i}\\
\end{array} \right)
\left( \begin{array}{c}
C^{EE}_{\ell,ij}\\
C^{EB}_{\ell,ij}\\
C^{BE}_{\ell,ij}\\
C^{BB}_{\ell,ij}\end{array} \right)\nonumber\\
&+&
(1+m_{0,i})
\left( \begin{array}{cccc}
 {\mathcal N}^{+,*}_{\ell,j}&{\mathcal N}^{-,*}_{\ell,j}&0&0\\
-{\mathcal N}^{-,*}_{\ell,j}&{\mathcal N}^{+,*}_{\ell,j}&0&0\\
0&0&{\mathcal N}^{+,*}_{\ell,j}&{\mathcal N}^{-,*}_{\ell,j}\\
0&0&-{\mathcal N}^{-,*}_{\ell,j}&{\mathcal N}^{+,*}_{\ell,j}\\
\end{array} \right)
\left( \begin{array}{c}
C^{EE}_{\ell,ij}\\
C^{EB}_{\ell,ij}\\
C^{BE}_{\ell,ij}\\
C^{BB}_{\ell,ij}\end{array} \right)\nonumber\\
&+&
(1+m_{0,j})\left( \begin{array}{c}
C^{c_E E}_{\ell,ij}\\
C^{c_E B}_{\ell,ij}\\
C^{c_B E}_{\ell,ij}\\
C^{c_B B}_{\ell,ij}\end{array} \right)
+
(1+m_{0,i})\left( \begin{array}{c}
C^{E c_E }_{\ell,ij}\\
C^{E c_B }_{\ell,ij}\\
C^{B c_E }_{\ell,ij}\\
C^{B c_B }_{\ell,ij}\end{array} \right)
+
\left( \begin{array}{c}
C^{c_E c_E}_{\ell,ij}\\
C^{c_E c_B}_{\ell,ij}\\
C^{c_B c_E}_{\ell,ij}\\
C^{c_B c_B}_{\ell,ij}\end{array} \right)\nonumber\\
&+&
\sum_{\ell'}
\left( \begin{array}{cccc}
 {\mathcal M}^{++}_{\ell\ell',ij}& {\mathcal M}^{+-}_{\ell\ell',ij}&
 {\mathcal M}^{-+}_{\ell\ell',ij}& {\mathcal M}^{--}_{\ell\ell',ij}\\
-{\mathcal M}^{+-}_{\ell\ell',ij}& {\mathcal M}^{++}_{\ell\ell',ij}&
-{\mathcal M}^{--}_{\ell\ell',ij}& {\mathcal M}^{-+}_{\ell\ell',ij}\\
-{\mathcal M}^{-+}_{\ell\ell',ij}&-{\mathcal M}^{--}_{\ell\ell',ij}&
 {\mathcal M}^{++}_{\ell\ell',ij}& {\mathcal M}^{+-}_{\ell\ell',ij}\\
 {\mathcal M}^{--}_{\ell\ell',ij}&-{\mathcal M}^{++}_{\ell\ell',ij}&
-{\mathcal M}^{+-}_{\ell\ell',ij}& {\mathcal M}^{++}_{\ell\ell',ij}\\
\end{array} \right)
\left( \begin{array}{c}
C^{EE}_{\ell',ij}\\
C^{EB}_{\ell',ij}\\
C^{BE}_{\ell',ij}\\
C^{BB}_{\ell',ij}\end{array} \right)\nonumber\\
&+&
\left( \begin{array}{c}
{\mathcal B}^{+EE}_{\ell\ell',ij}+({\mathcal B}^{+EE}_{\ell\ell',ji})^*+{\mathcal B}^{-BE}_{\ell\ell',ij}+({\mathcal B}^{-BE}_{\ell\ell',ji})^*\\
{\mathcal B}^{+EB}_{\ell\ell',ij}+({\mathcal B}^{+BE}_{\ell\ell',ji})^*+{\mathcal B}^{-BB}_{\ell\ell',ij}+({\mathcal B}^{-EE}_{\ell\ell',ji})^*\\
({\mathcal B}^{+EB}_{\ell\ell',ji})^*+{\mathcal B}^{+BE}_{\ell\ell',ij}+({\mathcal B}^{-BB}_{\ell\ell',ji})^*+{\mathcal B}^{-EE}_{\ell\ell',ij}\\
{\mathcal B}^{+BB}_{\ell\ell',ij}+({\mathcal B}^{+BB}_{\ell\ell',ji})^*-{\mathcal B}^{-EB}_{\ell\ell',ij}-({\mathcal B}^{-EB}_{\ell\ell',ji})^*  
\end{array} \right). 
\end{eqnarray}
The various matrices in the above expression are
\begin{eqnarray}
    {\mathcal M}^{XY}_{\ell\ell',ij}&=&\frac{1}{2\ell+1}\sum_{mm'} W^X_{\ell\ell'm m',i}(W^Y_{\ell\ell' m m',j})^*\nonumber\\
    {\mathcal N}^X_{\ell,i}&=&\frac{1}{2\ell+1}\sum_{m} W^X_{\ell\ell m m,i}\nonumber\\
    {\mathcal B}^{XGH}_{\ell\ell',ij}&=&\frac{1}{2\ell+1}\sum_{mm'} W^X_{\ell\ell'm m'_i}\gamma^G_{\ell'm',i} (c^H_{\ell m,j})^*, 
\end{eqnarray}
where $X=(+,-)$, $Y=(+,-)$, $G=(E, B)$ and $H=(E, B)$. In this case the linearised expressions, \extra{assuming no underlying B-modes, or EB power,} are 
\begin{eqnarray}
\widetilde C^{EE}_{\ell,ij}&\approx&
(1+m_{0,i}+m_{0,j})C^{EE}_{\ell,ij}+
\langle m^R_i\rangle C^{EE}_{\ell,ij}+
\langle m^R_j\rangle C^{EE}_{\ell,ij}+
C^{c_EE}_{\ell,ij}+
C^{Ec_E}_{\ell,ij},\nonumber\\
\widetilde C^{EB}_{\ell,ij}&\approx&
-\langle m^I_i\rangle C^{EE}_{\ell,ij}
+C^{Ec_B}_{\ell,ij},\nonumber\\
\widetilde C^{BE}_{\ell,ij}&\approx&
-\langle m^I_i\rangle C^{EE}_{\ell,ij}+
C^{c_BE}_{\ell,ij},\nonumber\\
\widetilde C^{BB}_{\ell,ij}&\approx&0.
\end{eqnarray}

\label{lastpage}
\end{document}